# Noncollinear spin textures and 90° domain walls in twisted XY magnets

Guanghui Cheng[1,2]†*, Shiva T. Konakanchi[3]†, Andres E. Llacsahuanga Allcca[4,5], Sanjeev Khare[3], Nithin Abraham[4,5], Yuqing Cao[8,9], Hechang Lei[8,9], Kenji Watanabe[10], Takashi Taniguchi[11], Pramey Upadhyaya[3,5]*, Yong P. Chen[1,3,4,5,6,7]*

[1]WPI Advanced Institute for Materials Research (AIMR), Tohoku University; Sendai 980-8577, Japan.

[2]Frontier Research Institute for Interdisciplinary Sciences, Tohoku University, 6-3 Aoba, Sendai 980-8578, Japan.

[3]Elmore Family School of Electrical and Computer Engineering, Purdue University, West Lafayette, Indiana 47907, USA.

[4]Department of Physics and Astronomy, Purdue University, West Lafayette, Indiana 47907, USA.

[5]Purdue Quantum Science and Engineering Institute and Birck Nanotechnology Center, Purdue University; West Lafayette, Indiana 47907, USA.

[6]Institute of Physics and Astronomy and Villum Centers for Hybrid Quantum Materials and devices, Aarhus University; 8000 Aarhus-C, Denmark.

[7]Institute for Materials Research (IMR), Tohoku University, Sendai 980-8577, Japan.

[8]School of Physics and Beijing Key Laboratory of Opto-electronic Functional Materials & Micro-nano Devices, Renmin University of China; Beijing 100872, China.

[9]Key Laboratory of Quantum State Construction and Manipulation (Ministry of Education), Renmin University of China, Beijing 100872, China.

[10]Research Center for Electronic and Optical Materials, National Institute for Materials Science; 1-1 Namiki, Tsukuba 305-0044, Japan.

[11]Research Center for Materials Nanoarchitectonics, National Institute for Materials Science; 1-1 Namiki, Tsukuba 305-0044, Japan.

†These authors contributed equally to this work.

*Corresponding authors. Emails: ghcheng16@gmail.com; prameyup@purdue.edu; yongchen@tohoku.ac.jp

## Abstract

**Twisted moiré magnets are promising in exploring noncollinear magnetic phases, yet current experimental studies have been restricted to uniaxial magnets, limiting the accessible phase space. Here, we demonstrate noncollinear moiré magnetism based on XY magnet $CrCl_3$. The tunneling magnetoconductance of twisted $CrCl_3$ exhibits multiple field-driven transitions in small-twist-angle devices, attributed to the coexisting antiferromagnetic and ferromagnetic domains with distinct susceptibilities. The inferred spin configuration depends on the layer number, reflecting the interlayer coupling strength between twisted layers. This moiré magnetism is remarkably robust, persisting up to twisted double nine-layer stacks. Combined with micromagnetic simulations, we identify the ground state as the predicted "twisted-s" phase featuring 90° domain walls. Finally, we demonstrate voltage control of these noncollinear phases, highlighting the electrically tunable twist-spintronics.**

Most magnetic materials discovered thus far, whether ferromagnetic (FM) or antiferromagnetic (AFM), possess collinear spins, which have dominated the field of spintronics over the past decades (*1*). However, their applications continue to face challenges, from slow spin dynamics to high switching energy. There has been a surge of interest in the noncollinear spin structures (*2, 3*), which not only circumvent these challenges, but also offer extra functionalities and nontrivial magnetic phases. For example, noncollinear spins enable novel spin transport (*4, 5*), unique forms of magnetoelectric coupling (*6, 7*), and can give rise to new classes of magnets, such as *p*-wave magnets (*8*). Therefore, considerable efforts have been devoted to searching for new noncollinear magnetic systems.

Recently, the introduction of moiré superlattices to the spin degree of freedom has created a new arena for exploring noncollinear magnetic phases (*9-20*). Due to the stacking-dependent interlayer exchange couplings (*21*), twisting leads to spatially varying stacking orders as well as magnetic orders. The coexistence and competition of distinct spin alignments within moiré supercells can give rise to noncollinear spin textures (*9-12*). To date, experimental studies of moiré magnetism remain largely confined to materials with strong uniaxial anisotropy (*13-19*). In the widely studied $CrI_3$ (*13-17*), the large easy-axis anisotropy keeps the moiré magnetism predominantly out-of-plane, so twisting mainly reshapes the competition between FM and AFM rather than unlocking the full manifold of spin configurations. As a result, moiré studies in easy-axis systems have accessed only a limited region of the magnetic phase space. Very recently, hexagonal-stacked twisted $CrI_3$ was shown to exhibit additional symmetry breaking and periodic in-plane spin textures (*22*), representing an important step beyond out-of-plane moiré order. However, these in-plane features remain limited and tied to domain-wall-like regions of an otherwise easy-axis background. This points naturally to the next frontier: twist engineering in a genuine XY magnet, where spins are intrinsically free to rotate within the plane (Fig. 1A).

XY magnetism was theoretically proposed in the 1970s (*23, 24*) and has been foundational to 2D phase transitions and many other topics in physics (*25*). Yet, material realizations of a genuine 2D XY magnet have only emerged very recently (*26*). The uncharted moiré magnetism based on XY magnets is expected to host a wealth of nontrivial magnetic phases (*10, 12, 20, 27*). Due to the absence of easy-axis anisotropy, XY magnets are particularly susceptible to moiré superlattices, leading to novel noncollinear spin structures such as the predicted twisted-s phase with 90° spin rotation (*10*), which may further support intriguing spin waves (*10, 11, 28*) and multiferroic behavior (*20*). Their swirling spin configurations can also stabilize topological spin textures (*12, 27, 29*), including merons and skyrmions. However, experimental identification of moiré magnetism in twisted XY magnets remains scarce, and the lack of experimental probes for in-plane spins in atomically thin materials further poses challenges.

In this work, we apply twist engineering to the van der Waals magnetic semiconductor $CrCl_3$, a genuinely easy-plane XY magnet (*26*), and obtain the first experimental signatures of moiré magnetism of XY magnet. As illustrated in Fig. 1A, all spins within a given layer are aligned by the intralayer FM coupling, and the interlayer magnetic coupling depends on the stacking (*12, 21*): monoclinic (M) stacking favors AFM coupling, while rhombohedral (R) stacking favors FM coupling. Natural few-layer $CrCl_3$ adopts monoclinic default stacking (*30*) and thus is a layered AFM. In twisted $CrCl_3$ (Fig. 1B), the interlayer atomic registry varies continuously, and the M and R stackings are the two locally stable structures possessing energy minima (*12, 27, 31*). Therefore, we focus on M and R stackings below.

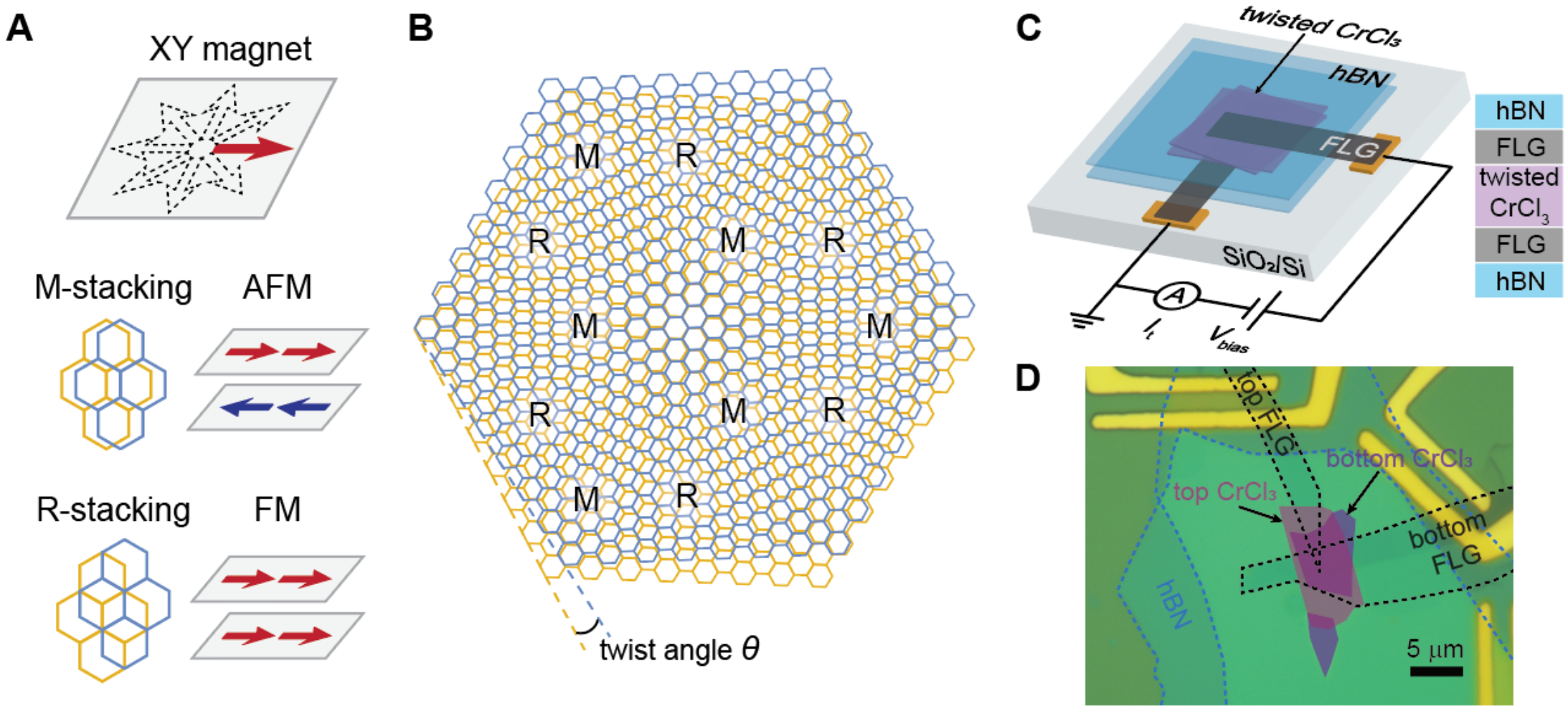

**Figure 1. Schematics of twisted $CrCl_3$ and magnetic tunnel junctions.** (**A**) XY spin model with continuous in-plane rotational symmetry, and one-to-one correspondence between M (R) stackings and interlayer AFM (FM) couplings. (**B**) Moiré superlattice of twisted $CrCl_3$ with a twist angle $\theta$. Only the hexagonal structures of Cr atoms are shown for simplicity. Regions of M and R stackings are labeled. (**C**) Schematic of a magnetic tunnel junction (MTJ) based on twisted $CrCl_3$ with an electrical measurement configuration. The purple, black, cyan and yellow blocks denote twisted $CrCl_3$, few-layer graphene (FLG), hexagonal boron nitride (hBN) and gold contacts, respectively. (**D**) Micrograph of a representative MTJ of 0.5°-twisted double four-layer $CrCl_3$. The stacking processes are shown in fig. S1. The top and bottom four-layer $CrCl_3$ are denoted by light and dark violet blocks, respectively. The two FLG flakes outlined by black dashed lines are in contact with the top and bottom surfaces of twisted $CrCl_3$. hBN flakes are outlined by cyan dashed lines.

**Magnetic tunnel junctions based on twisted $CrCl_3$**

Our previous studies involving $CrI_3$ with out-of-plane spins relied on the polar magneto-optical technique (*16, 32*), which is not suitable for in-plane spins in XY magnets. Here, we instead track the tunneling conductance in magnetic tunnel junctions (MTJs) with twisted $CrCl_3$ as the spin-filter barrier (Fig. 1, C and D), where the electron tunneling probability is determined by the interlayer spin alignment (*33*).

Figure 1D shows a representative junction of twisted $CrCl_3$ with a twist angle of $\theta = 0.5°$. Note that all $\theta$ in this work refer to the target twist angle during sample fabrication (*16*). Two few-layer graphene (FLG) flakes are in contact with the top and bottom surfaces of twisted $CrCl_3$ and serve as the tunneling electrodes. The overlapping region of the two FLG flakes corresponds to the tunneling area (~1.0 $\mu m^2$). The whole junction device is encapsulated by hexagonal boron nitride (hBN) flakes to avoid degradation. The tunneling conductance $G$ is obtained by applying a bias voltage $V_{\mathrm{bias}}$ and monitoring the tunneling current $I_t$, as illustrated in Fig. 1C. All measurements were conducted at 1.7 K unless otherwise specified (Materials and Methods).

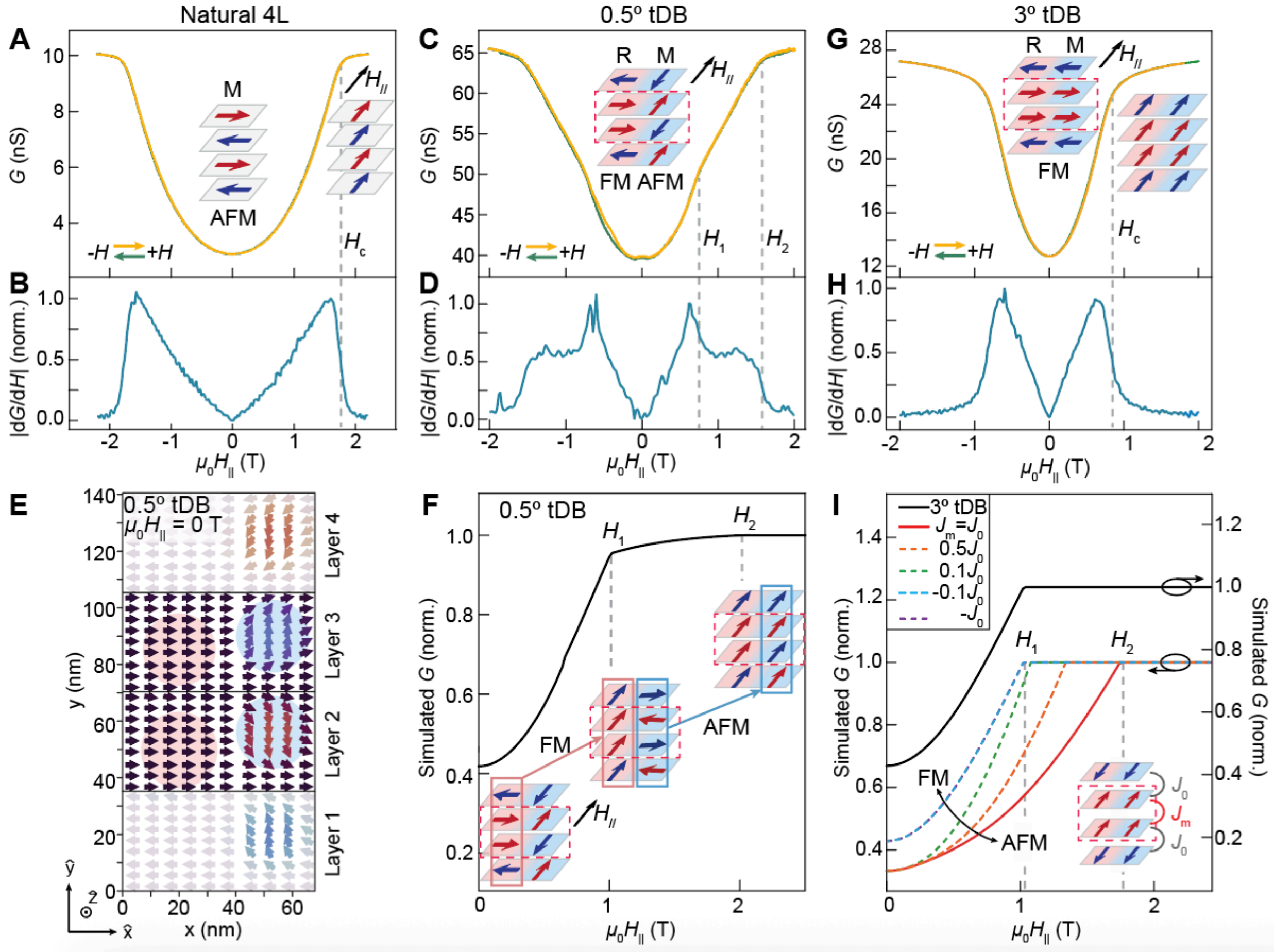

**Figure 2. Tunneling conductance of natural four-layer (4L) and twisted double bilayer (tDB) $CrCl_3$.** Magnetoconductance $G$(H) and its derivative $|dG/dH|$ of (**A** and **B**) natural 4L and (**C** and **D**) 0.5°-tDB $CrCl_3$ under an in-plane magnetic field $H_{\|}$ at a bias voltage of $V_{\mathrm{bias}} = 1$ V. The upward and downward sweeps of the field (yellow and green curves) almost coincide. Critical fields $H_c$ and $H_{1,2}$ are marked by dashed lines. Insets of (A): AFM ground state (left) and fully polarized state (right). Inset of (C): Ground state in the R and M stacking regions, with twisted layers highlighted by a red dashed rectangle. (**E**) Theoretically simulated noncollinear ground state of 0.5°-tDB $CrCl_3$ at zero field. The red and blue shaded areas correspond to the R (FM) and M (AFM) regions, respectively, between layers 2 and 3. Spins in layers 1 and 4 are assumed to be opposite (AFM coupled) to the corresponding spins in proximal layers 2 and 3, respectively. The full-scale moiré pattern is shown in fig. S4A. (**F**) Simulated $G$(H) of 0.5°-tDB $CrCl_3$ under $H_{\|}$. Two kinks emerge at $H_{1,2}$, representing the full polarization of the FM and AFM regions, respectively, denoted

by the insets. (**G** and **H**) $G$(H) and |d$G$/d$H$| of 3°-tDB $CrCl_3$ under $H_{\parallel}$ at $V_{bias}$ = 0.5 V. $H_c$ is marked by the dashed line. Insets of (G): Ground state in the R and M stacking regions (now both become FM), with twisted layers highlighted by a red dashed rectangle. (**I**) Simulated $G$(H), plotted on the left, of a 4L system with an averaged interlayer exchange $J_m$. $J_m$ varies from $J_0$ (AFM limit) to -$J_0$ (FM limit), leading to a change in the saturation field from $H_2$ to $H_1$ (marked by vertical dashed lines and consistent with theoretical values of $H_2 = 2H_J\cos^2\left(\frac{\pi}{8}\right) = 1.78$ T , and $H_1 = 2H_J\cos^2\left(\frac{\pi}{4}\right) = 1.04$ T , respectively, with $H_J = 1.04$ T , see also Materials and Methods). Simulated $G$(H) of a 3°-tDB is plotted on the right.

**Tunneling conductance of twisted double bilayers**

We first compare the results of natural four-layer (4L) and twisted double bilayer (tDB) $CrCl_3$ (2L plus 2L), as shown in Fig. 2, A to D. Current-voltage ($I$-$V$) curves of both cases are illustrated in fig. S2. Figure 2A shows the magnetoconductance $G(H)$ of natural 4L $CrCl_3$, which increases smoothly with an in-plane magnetic field $H_{\parallel}$ and saturates at the critical field $H_c$. At almost zero field, the interlayer-AFM-aligned spins undergo a spin-flop transition, pointing in the direction orthogonal to the external field (*34*) (left inset). When the field increases, the spins gradually cant toward the applied field until fully polarized (right inset). Only one phase transition is observed at $H_c$, exhibiting a sharp drop in the derivative |d$G$/d$H$| (Fig. 2B). Based on the linear-chain bond saturation model $H_c = 2H_J cos^2(\frac{\pi}{2N})$ (Materials and Methods), where $N$ is the number of layers (here 4), the extracted $H_c$ = 1.77 T corresponds to an interlayer exchange $H_J$ = 1.04 T, consistent with reported values of few-layer $CrCl_3$ (*30, 34*). Furthermore, $G(H)$ under a perpendicular magnetic field $H_{\perp}$ gives a larger critical field $H_{c\perp}$ =2.1 T (fig. S3, A and B), indicating the easy-plane anisotropy of natural $CrCl_3$ (*34*).

In sharp contrast, $G(H)$ of 0.5°-tDB $CrCl_3$ in Fig. 2C exhibits two kinks at $H_1$ (0.74 T) and $H_2$ (1.58 T), at which |d$G$/d$H$| in Fig. 2D shows pronounced drops. These observations indicate *two* phase transitions in tDB $CrCl_3$ and the presence of new magnetic orders in addition to the interlayer AFM order, which dominates in natural few-layer $CrCl_3$. $G(H)$ of 0.5°-tDB $CrCl_3$ under $H_{\perp}$ (which also exhibits two transitions) is shown in fig. S3, C and D. The larger perpendicular critical fields ($H_{1\perp}$ = 1.08 T and $H_{2\perp}$ = 1.9 T) suggest that the in-plane nature of magnetization persists in tDB $CrCl_3$. Therefore, only in-plane spin configurations and in-plane-field data are considered in the following. The difference between the perpendicular and in-plane critical fields reflects the magnetic anisotropy. $H_{1\perp} - H_1$ = 0.34 T and $H_{2\perp} - H_2$ = 0.32 T of the two transitions in tDB $CrCl_3$ are both close to $H_{c\perp} - H_c$ = 0.33 T in natural 4L $CrCl_3$, suggesting that the moiré superlattice can hardly change the magnetic anisotropy. Instead, it predominantly modulates the interlayer exchange coupling (*12, 21*).

To better understand the observations, we construct a micromagnetic model of tDB $CrCl_3$. Within the continuum limit (*10*), the free energy per unit area of the sample can be expressed as (*12, 16, 21*)

$$\mathcal{F} = \sum_{i=1}^{4} A(\nabla\vec{m}_i)^2 + J_0[\vec{m}_1 \cdot \vec{m}_2 + \Phi(\vec{r},\theta)\vec{m}_2 \cdot \vec{m}_3 + \vec{m}_3 \cdot \vec{m}_4] + \mathcal{F}_{\text{demag}}, \quad (1)$$

where $\vec{m}_i(r)$ is the position-dependent unit vector representing the spatial profile of local magnetization in the $i$th layer, $A$ is the intralayer exchange stiffness, $J_0 > 0$ is the AFM interlayer exchange coupling of natural $CrCl_3$ (*34*) and $\mathcal{F}_{\text{demag}}$ is the demagnetization energy arising from the dipole fields of the spins, which favors an in-plane arrangement of spins. $\Phi(\vec{r},\theta)$ represents a $\theta$-dependent spatial ($\vec{r}$) variation in the interlayer coupling between twisted layers (here layers 2 and 3) (*10-12*). No uniaxial anisotropy is included in our model for the easy-plane XY magnet (*30, 34*). The ground state is thus determined by a delicate balance between intralayer and interlayer exchange energies, with spatially varying modifications $\Phi(\vec{r},\theta)$ to the interlayer exchange.

Figure 2E shows the ground state configuration of 0.5°-tDB $CrCl_3$ (see the full moiré cell pattern in fig. S4A) at zero field. The magnetic moments in each layer lie within the sample plane. Between twisted layers 2-3, the R stacking regions are dominated by interlayer FM alignment, as reported in DFT calculations of $CrCl_3$ (*12*), while the M stacking regions are dominated by interlayer AFM alignment. The inset of Fig. 2C schematically shows the simplified ground state configuration. Notably, domain walls with 90° spin rotation between the FM regions and the AFM regions are present, suggesting that the sample attains the theoretically predicted noncollinear twisted-s phase (*10*). Between the untwisted layers 1-2 and 3-4, natural M stacking results in AFM order across the whole area. Therefore, the domains in the twisted layers are also imprinted onto the untwisted layers, and the system exhibits a four-domain-wall state for the parameters chosen in this simulation. The 90° spin rotation (compared to other possible angles) between the FM and AFM regions minimizes the domain walls' energy cost due to intralayer spin exchange stiffness. Such an emergent noncollinear twisted-s phase with 90° spin rotation is essentially enabled by the absence of uniaxial anisotropy in XY magnets. We note that the rotational XY symmetry is preserved in the ground state within the twisted-s phase in the absence of any symmetry-breaking fields.

Figure 2F shows the simulated $G(H)$ of 0.5°-tDB $CrCl_3$ obtained by adding the Zeeman term, $M_s \sum_{i=1}^{4} \vec{m}_i(r) \cdot \vec{H}$, to the free energy density in Eq. (1), with $M_s$ being the saturation magnetization. The conductance between any two adjacent layers is obtained by the formula $G_{ij} =$ $G_0 \sum_r (1 + p^2 \vec{m}_i(r) \cdot \vec{m}_j(r))$, where $G_0$ is the conductance when $\vec{m}_i(r) \cdot \vec{m}_j(r) = 0$ and $p$ is a dimensionless constant that depends on the spin polarization of the two magnetic layers at the Fermi level (*35, 36*). Notably, the simulated $G(H)$ shows two kinks at $H_1$ and $H_2$, consistent with the experimental observations in Fig. 2, C and D. This can be understood based on the field evolution of the two representative FM and AFM regions, as shown in the insets of Fig. 2F. At a vanishing small in-plane field, the field breaks the XY symmetry within the sample plane. Given that FM regions dominate over AFM regions (Fig. 2E), the spins in FM regions (denoted as FM spins) flop and point orthogonal to the external field to minimize the Zeeman energy of the entire four-layer system (where layers 1-2 and 3-4 remain AFM-coupled), while the spins in AFM regions (denoted as AFM spins) point along the external field. As the field is increased, the flopped FM spins cant toward the applied field and saturate at $H_1$ (i.e., fully polarized). Meanwhile, AFM spins maintain almost 90° domain walls with the FM spins and are nearly orthogonal to the field (see detailed field evolution of magnetization vectors in fig. S4B). Beyond $H_1$, the AFM spins start to cant toward the applied field and eventually saturate at $H_2$. These two processes are labeled FM and AFM, respectively, in the inset of Fig. 2F, and host different susceptibilities to the external field. As a result, two kinks are observed in $G(H)$. Carefully inspecting Fig. 2C and 2F reveals that the change in $G(H)$ from $H_1$ to $H_2$, which reflects the canting process of the AFM regions, is stronger in the experiment than in the simulation. This suggests that the DFT input parameters (*10-12*), upon which our micromagnetic simulations are built, may have underestimated the size or strength of the AFM regions, which can be further explored in future studies.

To better understand the saturation fields of the FM and AFM regions, we model the collinear order in either region using $J_m$, a uniform effective interlayer exchange, between the twisted layers and $J_0$, an AFM exchange of natural $CrCl_3$, between the remaining untwisted layers (inset of Fig. 2I). Figure 2I shows simulated $G$(H) as $J_m$ is tuned from $J_0$ (AFM limit) to $-J_0$ (FM limit). The saturation field lies near $H_2 = 2H_J \cos^2(\frac{\pi}{2N})$ for the AFM limit (same as $H_c$ of natural $CrCl_3$) and decreases toward $H_1 = 2H_J \cos^2(\frac{\pi}{N})$ as the system approaches the FM limit, where $N$ is the total number of layers (Materials and Methods, and $N = 4$ for Fig. 2I). This trend is consistent with the assignment of $H_1$ and $H_2$ to the FM and AFM domains, respectively, in small-twist-angle devices. The saturation field of an $N$-layer system depends on the ratio of the total number of AFM bonds that must be overcome by the external field to the total number of magnetic (Zeeman) layers. For an $N$-layer AFM system with interlayer exchange $J_0$, this ratio is $(N\text{-}1)/N$, and the saturation field is $H_2 = 2H_J \cos^2(\frac{\pi}{2N})$. When a uniform FM coupling is introduced between the middle two layers (irrespective of the strength of the FM coupling), this ratio becomes $(N\text{-}2)/N = (N/2\text{-}1)/(N/2)$, equivalent to an $N/2$-layer AFM system (assuming $N$ is an even number), with a lower saturation field $H_1 = 2H_J\cos^2(\frac{\pi}{N})$. As the coupling between the middle two layers is changed to AFM and its strength is raised from zero to $J_0$ in Fig. 2I, the critical saturation field increases from the FM case to the 4L AFM counterpart.

We further found that the moiré magnetism in twisted $CrCl_3$ can be controlled by the twist angle. As the twist angle increases, i.e., the moiré period $L$ decreases, the energy cost of forming domain walls (proportional to $L$) gradually surpasses the energy gain within the AFM or FM domains (proportional to $L^2$), ultimately causing the domain structures to disappear. Consistent with this picture, in 3°-tDB $CrCl_3$ (Fig. 2, G and H), only a single kink is observed at $H_c$ = 0.84 T, indicating the disappearance of the noncollinear domain structure and the restored uniform spin alignment within each layer. As illustrated in Fig. 2E, the majority of the area, including the R-stacking regions, is dominated by interlayer FM alignment and forms a "sea" that surrounds the AFM "bubble". Accordingly, the AFM bubbles are expected to shrink as the twist angle increases; eventually, the FM sea extends all across the layers, and the system transitions to a collinear FM order (even if the local interlayer exchange may still be modulated between FM and AFM based on local stacking orders). For 3°-tDB $CrCl_3$, the simulated $G$(H) (black curve of Fig. 2I) shows a critical field close to $H_1$, and the experimentally observed $H_c$ = 0.84 T (Fig. 2, G and H) is also near $H_1$. Together, these results verify that an interlayer FM order is established. The collinear FM ground state and its field-polarized state are illustrated in the insets of Fig. 2G.

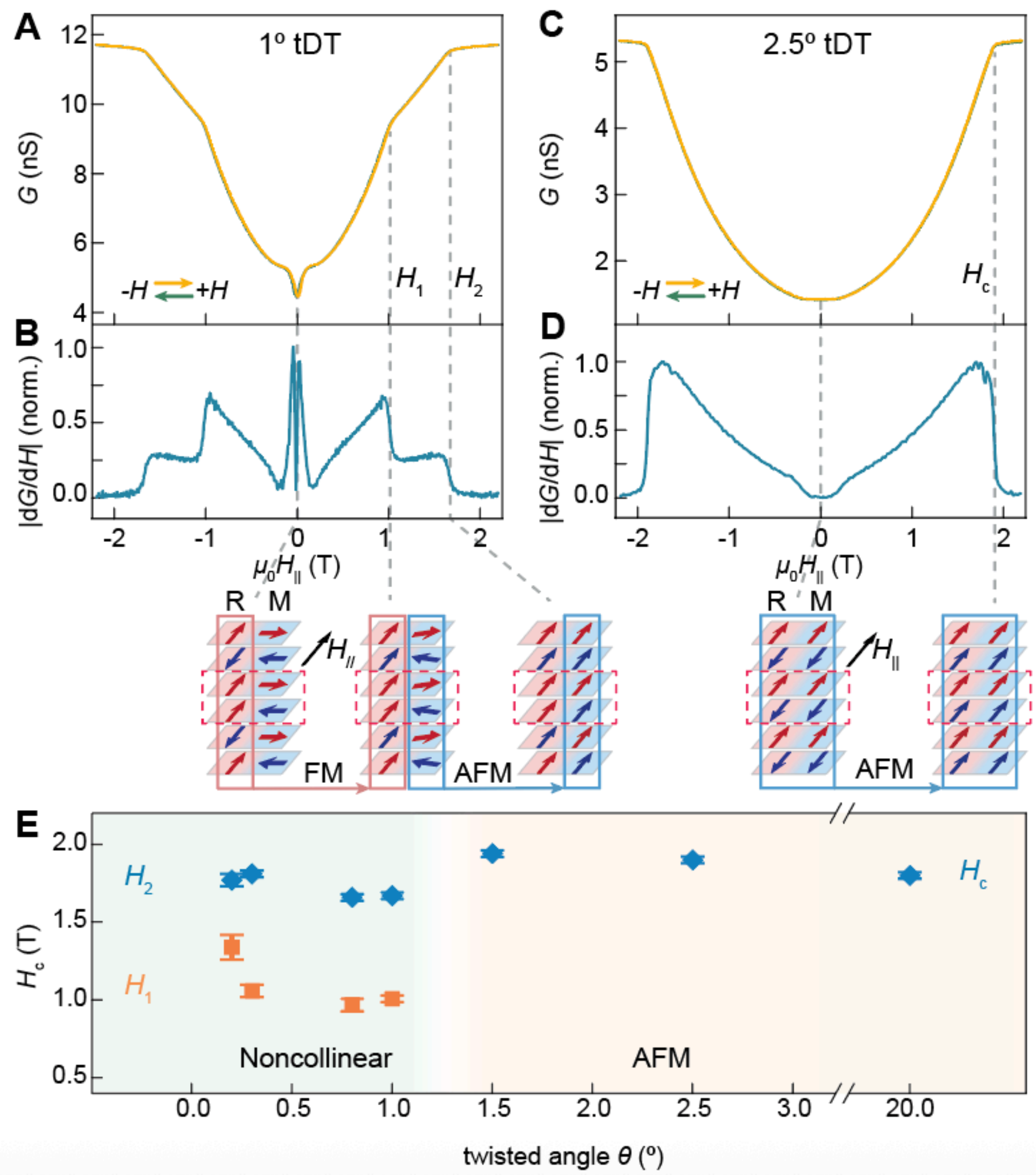

**Figure 3. Tunneling conductance of twisted double trilayer (tDT) $CrCl_3$.** $G$(H) and |d$G$/d$H$| of (**A** and **B**) 1°-tDT and (**C** and **D**) 2.5°-tDT $CrCl_3$ under $H_{\|}$ at $V_{bias}$ = 2.4 V. $H_{1,2}$ and $H_c$ are marked by dashed lines. Insets: Ground state configuration in the R and M stacking regions, with twisted layers highlighted by red dashed rectangles. The zero-field dip shown in Fig. 3 (**A** and **B**) is already observable in some natural few-layer samples (Ref. (*34*) and fig. S5) and is not the focus of this work. (**E**) Summarized critical fields as a function of the twist angle, extracted from tDT $CrCl_3$ devices with different twist angles. The noncollinear phase and collinear AFM phase dominating different twist angle regimes are marked by green and yellow shading, respectively.

**Tunneling conductance of twisted double trilayers**

Next, we study twisted double trilayer (tDT) $CrCl_3$ as a representative odd-plus-odd system. Figure 3, A and B, shows $G(H)$ and |d$G$/d$H$| of 1°-tDT $CrCl_3$ under $H_{\|}$. The presence of two kinks at $H_{1,2}$ suggests a noncollinear moiré phase (coexisting FM and AFM domains), similar to the tDB case. In addition, a zero-field dip emerges in $G(H)$. A similar feature at low-field is also observed sometimes in natural few-layer $CrCl_3$ (Ref. (*34*) and fig. S5) and may be attributed to the reduced interlayer exchange in some samples, as suggested by our simulations in fig. S6. Measurements under a perpendicular field (fig. S3, E and F) confirm the in-plane nature of the magnetization in tDT $CrCl_3$.

We study the zero-field ground state of tDT $CrCl_3$ using micromagnetic simulations (fig. S7A). Similar to tDB $CrCl_3$, the ground state of tDT $CrCl_3$ consists of coexisting FM and AFM domains separated by nearly 90° domain walls, i.e., a noncollinear twisted-s phase, as schematically illustrated below in Fig. 3B (leftmost configuration). Notably, unlike the tDB case, the FM domains in the tDT case exhibit a nonzero net magnetization. Consequently, at vanishingly small fields, the system orients itself such that the net magnetization, namely, the FM spins, aligns with the external field (in contrast to the tDB case in Fig. 2C, where the FM spins flop to be perpendicular to the field). Meanwhile, the AFM spins remain nearly orthogonal to the applied field. With increasing field, the sample experiences a spin-flop-like transition (*34, 37*), in which the FM spins flop away from (but not quite perpendicular to) the field direction (see fig. S7B for the detailed evolution). This behavior is caused by the increasing energy penalty of the opposite spins in other layers and interlayer exchange coupling. Upon further increasing the field, the FM spins (including those from the other untwisted, AFM-coupled layers) cant toward the applied field and saturate at $H_1$. Increasing the field beyond $H_1$ then drives the AFM spins to saturate at $H_2$ (labeled FM and AFM, respectively, below Fig. 3B).

For a larger-twist-angle device, 2.5°-tDT $CrCl_3$ (Fig. 3, C and D), only a single kink is observed at $H_c = 1.9$ T, indicating the absence of a noncollinear domain structure and the restored uniform spin alignment within each layer. We extract $H_{1,2}$ and $H_c$ from multiple tDT $CrCl_3$ devices with different twist angles and summarize them in Fig. 3E. The two-kink feature disappears for twist angles $\geq 1.5°$, signifying the transition from noncollinear to collinear magnetic order. Notably, different to the tDB case, $H_c$ in the large-twist-angle regime lies closer to $H_2$ than to $H_1$, suggesting that the large-twist-angle spin configuration in the tDT case follows AFM, rather than FM, alignment. This determination is further supported by simulations with tunable interlayer exchange (fig. S6). The AFM ground state and its field-polarized state are illustrated below Fig. 3D.

The AFM ground state in large-twist-angle tDT $CrCl_3$ is inconsistent with the DFT calculations performed for twisted 1L-plus-1L $CrCl_3$ (*12*), where FM domains dominate the majority of the area as an FM sea. In contrast, in tDT $CrCl_3$, our experimental results suggest that AFM domains may dominate, forming an AFM sea that surrounds FM bubbles. As the twist angle increases, the AFM sea extends throughout the system, and the system transitions to a collinear AFM order. This suggests that the dominant interlayer coupling between twisted layers may depend on the total layer number (FM sea in the tDB case versus AFM sea in the tDT case). An in-depth understanding requires further theoretical study.

As shown in fig. S8, the two-kink feature can be observed up to twisted double nine-layer $CrCl_3$ (an 18-layer system with only the middle two layers twisted, much thicker than other double-twisted/moiré systems studied thus far (*38*)), highlighting the robustness of moiré magnetism in $CrCl_3$.

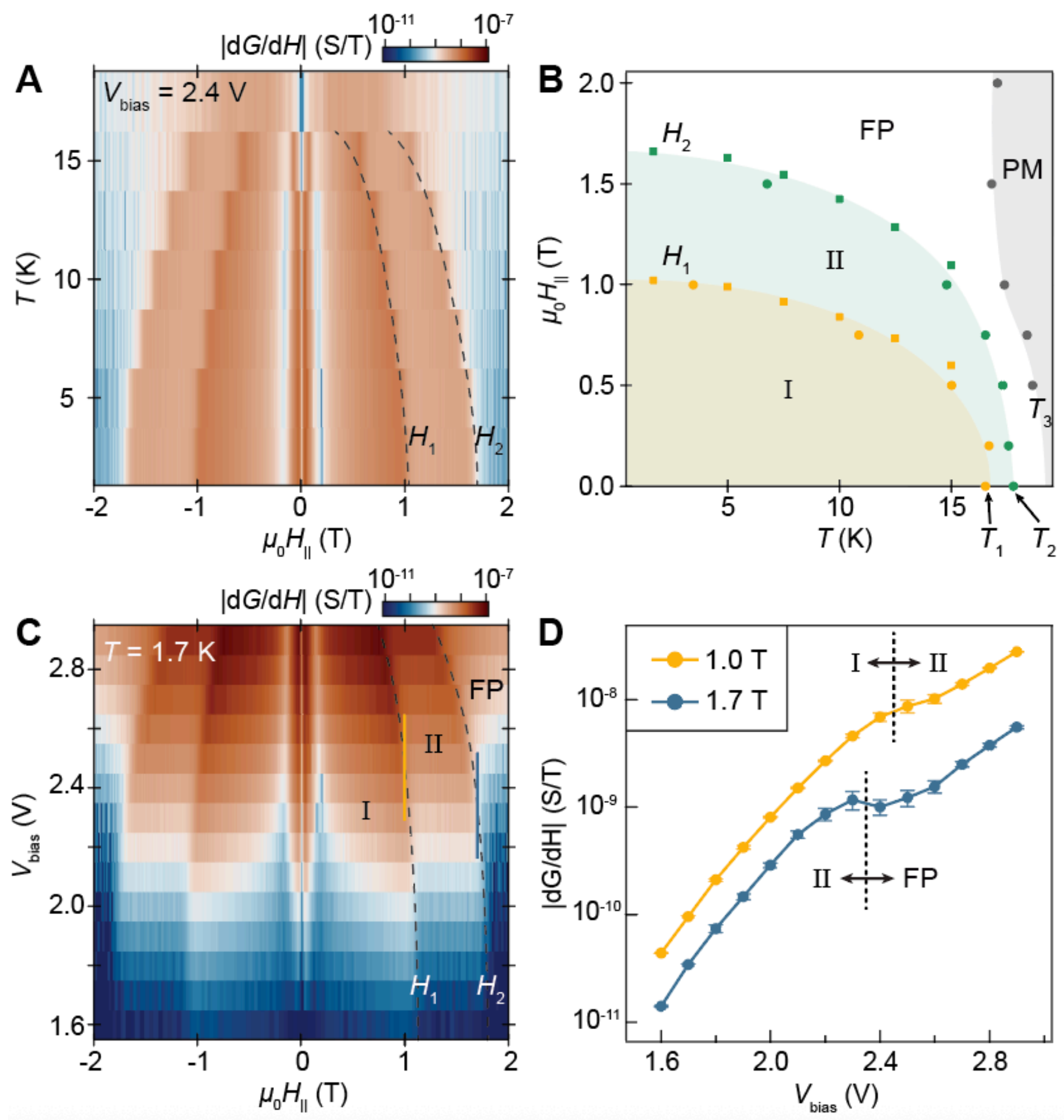

**Figure 4. Temperature dependence and bias control of the moiré magnetism in tDT $CrCl_3$.** (**A**) $|dG/dH|$ of 1°-tDT $CrCl_3$ plotted on a log scale as functions of $H_{\parallel}$ and temperature $T$ at $V_{bias}$ = 2.4 V. Temperature-dependent $G$(H) is shown in fig. S9. The critical fields are marked by dashed lines. (**B**) Phase diagram as functions of $T$ and $H_{\parallel}$. The rectangles and circles are extracted from $|dG/dH|$ in (A) and $dG/dT$ in fig. S10, respectively. Noncollinear phases I and II, fully polarized phase (FP), and paramagnetic phase (PM) are labeled. The phase boundaries of I-II and II-FP,

determined by $H_1$ and $H_2$, intersect the horizontal axis at $T_1$ and $T_2$, respectively. $T_3$ represents the PM phase transition temperature. (**C**) |d$G$/d$H$| of 1°-tDT $CrCl_3$ plotted on a log scale as functions of $H_{\parallel}$ and $V_{bias}$. Bias-dependent $G$(H) is shown in fig. S11. The critical fields are marked by dashed lines. (**D**) |d$G$/d$H$| as a function of $V_{bias}$ at fixed $H_{\parallel}$ of 1.0 T and 1.7 T, respectively. The kinks, marked by the dashed lines, suggest bias-induced phase transitions I-II and II-FP, respectively. These bias-induced phase transitions are also illustrated in (C) as yellow and blue solid lines.

**Temperature dependence and bias control of moiré magnetism**

Figure 4A shows the temperature dependence of |d$G$/d$H$| in 1°-tDT $CrCl_3$. The sharp changes in the color plot mark the magnetic transitions at critical fields $H_{1,2}$ (dashed lines). As the temperature is increased, all critical fields shift to lower values due to the quenching of magnetic exchange interactions by thermal energy. These critical fields are extracted and plotted in Fig. 4B as rectangles. We further obtain the critical temperatures at different fields through the temperature derivative d$G$/d$T$ (fig. S10), plotted in Fig. 4B as circles. These two sets of critical points reflect the same phase transitions and together determine the boundaries of the phase diagram in Fig. 4B. As discussed above, the transitions at $H_{1,2}$ are attributed to the saturation processes of the FM and AFM domains, respectively. The noncollinear phases are labeled I and II. The fully polarized phase (FP) and paramagnetic phase (PM) are located at high field and high temperature, respectively. Notably, the phase boundaries of I-II and II-FP, determined by $H_1$ and $H_2$, intersect the horizontal axis at different temperatures, $T_1$ = 16.5 K and $T_2$ = 17.8 K, respectively. Compared with the Neel temperature $T_N$ ~ 18.6 K of natural 6L $CrCl_3$ (fig. S12), the closer $T_2$ corroborates the association of the boundary $H_2$ with the AFM domains. In the temperature range $T_1<T<T_2$, only a single field-induced transition is observed at $H_2$, suggesting that the thermal excitation suppresses the noncollinear domain structure and restores the uniform AFM arrangement.

Achieving electrical control of magnetism is essential for developing spintronic devices. Here, we employ the bias voltage $V_{bias}$, a readily accessible knob in MTJs (*39, 40*), to modulate moiré magnetism in twisted $CrCl_3$, as shown in Figure 4, C and D. The color plot of |d$G$/d$H$| is shown as functions of $H_{\parallel}$ and $V_{bias}$. $V_{bias}$ is kept low (< 3 V) given that the bias-induced heat effect starts to play a role and shift the Neel temperature above $V_{bias}$ = 3.2 V, as shown in fig. S13. Figure 4C shows that $H_{1,2}$ significantly shift to lower fields when $V_{bias}$ increases from 1.6 V to 2.8 V, indicating an overall decrease in interlayer exchange coupling. These behaviors in response to $V_{bias}$ may suggest an intriguing magnetoelectric effect in twisted $CrCl_3$. Essentially, $V_{bias}$ can be regarded as a vertical electric field applied across twisted $CrCl_3$ (e.g., $V_{bias}$ = 2.4 V corresponds to a significant vertical electric field of ~ 0.67 V/nm). The interlayer twisting between $CrCl_3$ layers can break the lattice inversion symmetry (*41*), which allows for a direct electric-field modification of the interlayer exchange interactions via spin-charge coupling (*32*).

We further demonstrate bias-induced phase transitions in Fig. 4D. With the magnetic field fixed at 1.0 T and 1.7 T, $|dG/dH|$ shows remarkable kinks in both cases as $V_{bias}$ increases, marked by the dashed lines. The two kinks correspond to the phase transitions I-II and II-FP, which are also illustrated in Fig. 4C as yellow and blue solid lines, respectively. Such bias-induced switching between magnetic phases demonstrates an electrical method to control the noncollinear moiré phases in twisted $CrCl_3$.

**Conclusions**

In conclusion, we report the observation of noncollinear moiré magnetism, in particular a twisted-s phase featuring 90° domain walls, in a twisted XY magnet. The observed multiple phase transitions and their evolution with magnetic field, temperature and bias voltage establish comprehensive control over the noncollinear phases. The revealed 90° domain walls offer unique functionality due to their distinct anisotropy, magnetostatic, and spin-transport signatures, and can find applications in energy-efficient spintronic memory and logic devices. The easy-plane XY spins, combined with the twist degree of freedom, facilitate a rich variety of nontrivial spin textures and may host topological quasiparticles (*12, 27*) and spin superfluidity (*42, 43*). Moreover, symmetry engineering in the twisted XY system by heterostructures or substrates may further create nontrivial moiré potentials, realizing, for example, $q$-state clock physics (*44*), which favors discrete in-plane angles and allows topological transitions beyond the conventional Kosterlitz-Thouless scenario.

**ACKNOWLEDGMENTS**

S.T.K. and P.U. would like to thank Mohammad Mushfiqur Rahman for invaluable help with developing the micromagnetic codes. G.H.C. further acknowledges experimental assistance from H. Zhang and C. Zeng. **Funding:** G.H.C. and Y.P.C acknowledge partial support from JSPS KAKENHI grants (24K16998, 22H00278, 25H00413 and 25K22195), FRIS Research Funds and AIMR, Tohoku University. S.T.K., S.K., A.L.A., N.A., Y.P.C. and P.U. also acknowledge support from the Quantum Science Center (QSC). K.W. and T.T. acknowledge support from the JSPS KAKENHI (21H05233 and 23H02052), the CREST (JPMJCR24A5), JST and World Premier International Research Center Initiative (WPI), MEXT, Japan. H.C.L. was supported by the National Key R&D Program of China (2023YFA1406500 and 2022YFA1403800) and the National Natural Science Foundation of China (12274459). **Author contributions:** G.H.C. and Y.P.C. conceived the project. G.H.C. fabricated the devices and performed experiments, assisted by A.E.L.A. and N.A. S.T.K., S.K. and P.U. performed supporting theoretical analysis. Y.Q.C. and H.C.L. provided bulk $CrCl_3$ crystals. K.W. and T.T. provided bulk hBN crystals. Y.P.C. and P.U. supervised the project. G.H.C., S.T.K., P.U. and Y.P.C. wrote the manuscript with input from all co-authors. **Competing interests**: The authors declare no competing financial interests. **Data, code, and materials availability:** Data are available in the manuscript or supplementary materials. Additional data are available from the authors upon reasonable request.

**SUPPLEMENTARY MATERIALS**

Materials and Methods

Figs. S1 to S13

## REFERENCES AND NOTES

1. I. Žutić, J. Fabian, S. Das Sarma, Spintronics: Fundamentals and applications. *Rev. Mod. Phys.* **76**, 323-410 (2004).
2. B. H. Rimmler, B. Pal, S. S. P. Parkin, Non-collinear antiferromagnetic spintronics. *Nat. Rev. Mater.* **10**, 109-127 (2025).
3. A. Brataas, G. E. W. Bauer, P. J. Kelly, Non-collinear magnetoelectronics. *Phys. Rep.* **427**, 157-255 (2006).
4. S. Nakatsuji, N. Kiyohara, T. Higo, Large anomalous Hall effect in a non-collinear antiferromagnet at room temperature. *Nature* **527**, 212-215 (2015).
5. H. Chen, Q. Niu, A. H. MacDonald, Anomalous Hall effect arising from noncollinear antiferromagnetism. *Phys. Rev. Lett.* **112**, 017205 (2014).
6. T. Kimura, Spiral magnets as magnetoelectrics. *Annu. Rev. Mater. Res.* **37**, 387-413 (2007).
7. H. Katsura, N. Nagaosa, A. V. Balatsky, Spin current and magnetoelectric effect in noncollinear magnets. *Phys. Rev. Lett.* **95**, 057205 (2005).
8. A. Birk Hellenes *et al.*, P-wave magnets. 2023 (10.48550/arXiv.2309.01607).
9. Q. J. Tong, F. Liu, J. Xiao, W. Yao, Skyrmions in the moiré of van der Waals 2D magnets. *Nano Lett.* **18**, 7194-7199 (2018).
10. K. Hejazi, Z. X. Luo, L. Balents, Noncollinear phases in moiré magnets. *Proc. Natl. Acad. Sci. U.S.A.* **117**, 10721-10726 (2020).
11. C. Wang, Y. Gao, H. Y. Lv, X. D. Xu, D. Xiao, Stacking domain wall magnons in twisted van der Waals magnets. *Phys. Rev. Lett.* **125**, 247201 (2020).
12. M. Akram *et al.*, Moiré skyrmions and chiral magnetic phases in twisted $CrX_3$ (X = I, Br, and Cl) bilayers. *Nano Lett.* **21**, 6633-6639 (2021).
13. T. Song *et al.*, Direct visualization of magnetic domains and moiré magnetism in twisted 2D magnets. *Science* **374**, 1140-1144 (2021).
14. Y. Xu *et al.*, Coexisting ferromagnetic–antiferromagnetic state in twisted bilayer $CrI_3$. *Nat. Nanotechnol.* **17**, 143-147 (2021).
15. H. Xie *et al.*, Twist engineering of the two-dimensional magnetism in double bilayer chromium triiodide homostructures. *Nat. Phys.* **18**, 30–36 (2021).
16. G. Cheng *et al.*, Electrically tunable moiré magnetism in twisted double bilayers of chromium triiodide. *Nat. Electron.* **6**, 434-442 (2023).
17. B. Yang *et al.*, Macroscopic tunneling probe of moiré spin textures in twisted $CrI_3$. *Nat. Commun.* **15**, 4982 (2024).
18. Y. Chen *et al.*, Twist-assisted all-antiferromagnetic tunnel junction in the atomic limit. *Nature* **632**, 1045-1051 (2024).
19. F. Yao *et al.*, Moiré magnetism in $CrBr_3$ multilayers emerging from differential strain. *Nat. Commun.* **15**, 10377 (2024).
20. A. O. Fumega, J. L. Lado, Moiré-driven multiferroic order in twisted $CrCl_3$, $CrBr_3$ and $CrI_3$ bilayers. *2D Mater.* **10**, 025026 (2023).
21. N. Sivadas, S. Okamoto, X. D. Xu, C. J. Fennie, D. Xiao, Stacking-dependent magnetism in bilayer $CrI_3$. *Nano Lett.* **18**, 7658-7664 (2018).
22. Z. Sun *et al.*, Tunable symmetry breaking in a hexagonal-stacked moiré magnet. *Nat. Phys.*, (2026).
23. J. M. Kosterlitz, D. J. Thouless, Ordering, metastability and phase transitions in two-dimensional systems. *J. Phys. C: Solid State Phys.* **6**, 1181 (1973).
24. J. M. Kosterlitz, The critical properties of the two-dimensional xy model. *J. Phys. C: Solid State Phys.* **7**, 1046 (1974).
25. J. M. Kosterlitz, Nobel Lecture: Topological defects and phase transitions. *Rev. Mod. Phys.* **89**, 040501 (2017).
26. A. Bedoya-Pinto *et al.*, Intrinsic 2D-XY ferromagnetism in a van der Waals monolayer. *Science* **374**, 616-620 (2021).

27. K.-M. Kim, G. Go, M. J. Park, S. K. Kim, Emergence of stable meron quartets in twisted magnets. *Nano Lett.* **24**, 74-81 (2024).
28. M. M. Rahman, A. Rustagi, Y. Tserkovnyak, P. Upadhyaya, Electrically active domain wall magnons in layered van der Waals antiferromagnets. *Phys. Rev. Lett.* **130**, 036701 (2023).
29. X. Lu, R. Fei, L. Zhu, L. Yang, Meron-like topological spin defects in monolayer $CrCl_3$. *Nat. Commun.* **11**, 4724 (2020).
30. D. R. Klein *et al.*, Enhancement of interlayer exchange in an ultrathin two-dimensional magnet. *Nat. Phys.* **15**, 1255-1260 (2019).
31. F. Yao *et al.*, Multiple antiferromagnetic phases and magnetic anisotropy in exfoliated $CrBr_3$ multilayers. *Nat. Commun.* **14**, 4969 (2023).
32. G. Cheng *et al.*, Emergence of electric-field-tunable interfacial ferromagnetism in 2D antiferromagnet heterostructures. *Nat. Commun.* **13**, 7348 (2022).
33. D. R. Klein *et al.*, Probing magnetism in 2D van der Waals crystalline insulators via electron tunneling. *Science* **360**, 1218-1222 (2018).
34. Z. Wang *et al.*, Determining the phase diagram of atomically thin layered antiferromagnet $CrCl_3$. *Nat. Nanotechnol.* **14**, 1116-1122 (2019).
35. J. C. Slonczewski, Conductance and exchange coupling of two ferromagnets separated by a tunneling barrier. *Phys. Rev. B* **39**, 6995-7002 (1989).
36. R. H. Fowler, L. Nordheim, Electron emission in intense electric fields. *Proc. R. Soc. Lond. A* **119**, 173-181 (1997).
37. A. N. Bogdanov, A. V. Zhuravlev, U. K. Rößler, Spin-flop transition in uniaxial antiferromagnets: Magnetic phases, reorientation effects, and multidomain states. *Phys. Rev. B* **75**, 094425 (2007).
38. D. Waters *et al.*, Topological flat bands in a family of multilayer graphene moiré lattices. *Nat. Commun.* **15**, 10552 (2024).
39. G.-X. Miao, M. Müller, J. S. Moodera, Magnetoresistance in double spin filter tunnel junctions with nonmagnetic electrodes and its unconventional bias dependence. *Phys. Rev. Lett.* **102**, 076601 (2009).
40. S.-C. Oh *et al.*, Bias-voltage dependence of perpendicular spin-transfer torque in asymmetric MgO-based magnetic tunnel junctions. *Nat. Phys.* **5**, 898-902 (2009).
41. L. Du *et al.*, Engineering symmetry breaking in 2D layered materials. *Nat. Rev. Phys.* **3**, 193-206 (2021).
42. J. König, M. C. Bønsager, A. H. MacDonald, Dissipationless spin transport in thin film ferromagnets. *Phys. Rev. Lett.* **87**, 187202 (2001).
43. S. Takei, Y. Tserkovnyak, Superfluid spin transport through easy-plane ferromagnetic insulators. *Phys. Rev. Lett.* **112**, 227201 (2014).
44. Z.-Q. Li *et al.*, Critical properties of the two-dimensional $q$-state clock model. *Phys. Rev. E* **101**, 060105 (2020).